\documentclass[aps,amssymb,floatfix,prd,amsmath,preprintnumbers]{revtex4}
\usepackage{epstopdf}
\usepackage{capt-of}
\usepackage{graphicx}  
\usepackage{dcolumn}   
\usepackage{bm}
\usepackage[font=scriptsize]{caption}
\usepackage{mwe}
\begin{document}
\input epsf.tex
\title{\textbf{Anisotropic cosmological models with two fluids}}
\author{B. Mishra\footnote{
Department of Mathematics, Birla Institute of Technology and Science-Pilani,
Hyderabad Campus, Hyderabad-500078, India, email: bivudutta@yahoo.com},
Pratik P. Ray \footnote{
Department of Mathematics, Birla Institute of Technology and Science-Pilani,
Hyderabad Campus, Hyderabad-500078, India, email: pratik.chika9876@gmail.com}
, S.K.J. Pacif \footnote{
Centre of Theoretical Physics, Jamia Millia Islamia, New Delhi-110025,
India, email: shibesh.math@gmail.com} }
\affiliation{ }

\begin{abstract}
\begin{center}
\textbf{Abstract}
\end{center}

In this paper, anisotropic dark energy cosmological models have been constructed in a Bianchi-V space-time where the energy momentum tensor consisting of two non-interacting fluids namely bulk viscous fluid and dark energy fluid. Two different models are constructed based on the power law cosmology and de Sitter universe. The constructed model also embedded with different pressure gradients along different spatial directions.  The variable equation of state (EoS) parameter, skewness parameters for both the models are obtained and analysed. The physical properties of the models obtained with the use of scale factors of power law and de Sitter law are also presented.
\end{abstract}
\maketitle

\address{Department of Mathematics,\\
Birla Institute of Technology and Science-Pilani, Hyderabad Campus,\\
Hyderabad-500078, India\\
bivudutta@yahoo.com, pratik.chika9876@gmail.com}

\input epsf.tex 


\section{Introduction}

The most popular problem in modern cosmology has been invoked by the current discovery of accelerated expansion of the universe. This has been confirmed as an established fact through different observational data, such as Type Ia Supernova(SNIa)\citep{riess, perm}, CMB radiation \cite{A, B, C}, gravitational lensing \cite{D, E} etc. This development is explained at the backdrop of general relativity (GR)through the introduction of an unknown energy source termed as dark energy (DE). This DE provides a repulsive gravity that helps in driving the acceleration by generating a strong negative force leading to an anti gravity effect. In Friedman-Robertson-Walker (FRW) universes, viscosity appears as the only dissipate phenomena, so a considerable amount of interest is seen in the study of cosmological models with bulk viscous fluid. In the inflationary phase, the contribution of bulk viscosity is well recognized which gives rise to a negative pressure that simulates a repulsive gravity. The equation of
state (EoS) parameter of the viscous fluid having  value lower than -1 generally considered to be 
significant in the context of DE cosmology. Observational results already indicated that the value of $\omega \ll -1$ \cite{A, B, C}, however DE crossing phantom divide line having $\omega \geq -1$ is lightly favoured. In consistent with the observational results, Copeland et al. \cite {cope}, Li et al. \cite{li} have used the scalar field approach with an introduced of time dependent EoS parameter to obtain the acceptable range for $\omega$. Another way to achieve this result is to reveal the solutions of the Einstein's field equations by incorporating some kinematical assumptions, which are in consistent with the observed kinematics of the Universe. As a testimony to this, Hubble parameter has been widely used to obtain explicit accelerating cosmological models in the framework of spatially homogeneous space-time\cite{kumar1}.\\

It can be noted that dominance of an anisotropic stress gives rise to an anisotropic expansion. This dominance will have a considerable impact via anisotropic stress on cosmological evolutions such as magnetic fields, hydrodynamic shear viscosity, collision less relativistic particles etc. \citep{barrow1,barrow2}. However, researches on DE with homogeneous and anisotropic space-time with time varying EoS parameter observed that at late time of cosmic evolution, DE yields isotropic pressure \cite {F,G}. On the other hand, several researchers \cite{cruz, hoftu, bennet1} focused on the fact that Wilkinson Microwave Anisotropy Probe (WMAP) data \citep{spergel,komat1} requires Bianchi type morphology instead of Friedman-Robertson-Walker (FRW) type for better accurate explanation of the anisotropic universe. Campanelli \citep{campa4}, revealed that irrespective of the level of anisotropy in
geometry of the universe and dark energy EoS, the SNIa data are always more consistent with standard isotropic universe. \\

Mishra et al. \cite{mishra1} have constructed the cosmological model based on pressure anisotropy in the presence of a gauge function whereas Mishra et al. \citep{mishra17} have studied the anisotropic universe with general forms of scale factor. Several cosmological models were obtained with constant deceleration parameter where the matter is in the form of perfect fluid or ordinary matter. However, many of those matters are not enough to describe the dynamics of an accelerating universe relating to anisotropy. This motivates us to consider the model of the accelerating universe filled with non-interacting fluids \cite{tripa1, ppr1, ppr2}. Akarsu and Kilinc \citep{akarsu3, akarsu4} have assumed constant deceleration parameter to construct and investigate DE models in Bianchi I and Bianchi III space-time. Yadav et al.\cite{yadav1} have assumed variable EoS parameter but constant deceleration parameter to construct the DE cosmological model in a locally rotationally symmetric Bianchi V space-time. Theoretical models of interacting and non-interacting DE have been discussed widely in the literature \cite{amir1, pradhan1, shey}. The paper is  arranged as follows: in section II, a mathematical formalism of an anisotropic DE universe is presented along with the relevant physical parameters. Two dark energy cosmological models one with power law cosmology and the other one with de Sitter universe has been constructed and analysed in section III. The summary is given in section IV.\\
\section{Formalism}
From an observational viewpoint, one of the most important result is the theorem of Wald \cite{wald84}, which states that universe with accelerating expansion tends towards isotropy at late phase. As a matter of fact, if the universe undergoes an early period of inflation, the present day universe will seem to be highly isotropic. Further, since the universe has now started accelerating, any kind of anisotropy will remain small in the late phase of cosmic acceleration. Bianchi universes are the class of cosmological models that are homogeneous but not necessarily isotropic on spatial slices. It contains, as a subclass, the standard isotropic model known as FRW universe. Calculations of nucleosynthesis and microwave background anisotropies in Bianchi models have been compared against data from the real Universe, typically gives null results which can be translated into upper limits on anisotropy. Tentative detections of non-zero anisotropic shear by Jaffe et al. \cite{jaffe05} are currently believed to be in consistent with other known cosmological parameters \cite{planck15} and with polarization of the microwave background \cite{pont07}. However these models remain widely-studied for their pedagogical value, mainly, making them tractable exact solutions of Einstein's field equation.\\

In the present paper, we are interested to study about the behaviour of anisotropy universe in the DE cosmological model. The standard FRW universe is homogeneous and isotropic. But in order to address the small scale anisotropy nature of the universe, Bianchi space-time is well accepted as it represents a globally hyperbolic spatially homogeneous, but not isotropic space-time. Among all 9 space-times of Bianchi, Bianchi V space-time is very intuitive as it has more degrees of freedom characterized by Lie groups and generates pseudo spherical space. Hence, in order to construct an anisotropic DE cosmological model in GR, we have considered here Bianchi type V space-time in the form, 
\begin{equation}  \label{eq1}
ds^{2}= dt^{2}-\sum_{i=1}^{3}e^{2a_{i}x}b_{i}^{2} dx_{i}^{2}
\end{equation}

where, $b_{i}=b_i(t),i=1,2,3$ are the directional scale factors considered to be different along three orthogonal directions and thereby provide a source for anisotropic expansion. Here, we choose, $a_1=0,$ $a_{2}=a_{3}=a,$ $a$ being a non zero arbitrary constant. Assuming GR is well defined at cosmic scales, we incorporate Einstein's field equations, 
\begin{equation}  \label{eq2}
G_{ij}\cong R_{ij}-\frac{1}{2}Rg_{ij}=\kappa T_{ij},
\end{equation}

where, $G_{ij}$, $R_{ij}$, $R$ and $T_{ij}$ respectively denotes the Einstein tensor, Ricci tensor, Ricci scalar and  total effective energy momentum tensor(EMT) and $\kappa=\frac{8\pi G}{c^4}$. $G$ is the Newtonian gravitational constant and $c$ is the speed of light with $8\pi G=c=1$. Here, EMT consists of two different components; the barotropic bulk viscous fluid $(T_{ij}^{vis})$ and DE fluid $(T_{ij}^{de})$. In case of barotropic cosmic fluid, the proper pressure $p$ is given as, $p= \xi \rho,$ $(0 \leq \xi \leq 1)$. The pressure with the contribution from bulk viscosity is also directly proportional to energy density, i.e, $3 \zeta H= \epsilon_{0} \rho$, where $\epsilon_{0} \geq 0$ is the proportionality constant \cite{tripathy10, brevik} and $\zeta u^i_{;i}=3\zeta H$. Hence, the effect of both proper pressure and barotropic bulk viscous pressure together can be expressed as,
\begin{align*}
\bar{p} = p-3 \zeta H = (\xi - \epsilon_{0}) \rho= \epsilon \rho
\end{align*}

where $\epsilon$ is the bulk viscous coefficient. One can infer that such a relation bears similarity to the pressure term with the contribution from perfect fluid ($p=\omega \rho$), where $\omega$ is the equation of state parameter for perfect fluid. However, a major part of the EoS in $\bar p$ of the present model comes from barotropic bulk viscosity. So, in no viscosity condition, the pressure term $\bar p$,  reduces to the pressure of perfect fluid. Hence, the EMT for viscous fluid is given as,

\begin{align}  \label{eq3}
T_{ij}^{vis}= (\rho + \bar{p}) u_{i}u_{j}- \bar{p} g_{ij}
\end{align}

where, $u^{i}$ is the four velocity vectors of the fluid. It may be noted that there is no observational reasons to conclude that pressure is isotropic in DE. However, since the fluids are co-moving, one may get this isotropic pressure in DE. Subsequently both the DE fluid and EoS parameter are direction dependent. Hence, the EMT of DE fluid is considered in the form,

\begin{align}  \label{eq4}
T_{ij}^{de} &= diag[\rho^{de}, -p^{de}_{x}, -p^{de}_{y}, -p^{de}_{z}]  \notag
\\
&= diag[\rho, -\omega^{de}_{x}, -\omega^{de}_{y}, -\omega^{de}_{z}] \rho^{de}
\notag \\
&= diag[1, -(\omega^{de}+ \delta), -(\omega^{de}+ \gamma), -(\omega^{de}+
\eta)]\rho^{de}
\end{align}

where, $\omega^{de}$ is the EoS parameter of the DE fluid along the dimensional axis $x$,$y$, and $z$. $\rho^{de}$ is the dark energy density. The deviations of $\omega^{de}$ from $x$,$y$, and $z$ axes respectively denotes the skewness parameters $\delta$, $\gamma$ and $\eta$. In the presence of EMT, Einstein's field equations (2) corresponding to Bianchi type V space-time (1) lead to the following:

\begin{align}
& \frac{\ddot{b_{2}}}{b_{2}}+\frac{\ddot{b_{3}}}{b_{3}}+\frac{\dot{b_{2}}%
\dot{b_{3}}}{b_{2}b_{3}}-\frac{a^{2}}{b_{1}^{2}}=-p+3 \zeta
H-(\omega^{de}+\delta)\rho^{de}  \label{eq5} \\
& \frac{\ddot{b_{1}}}{b_{1}}+\frac{\ddot{b_{3}}}{b_{3}}+\frac{\dot{b_{1}}%
\dot{b_{3}}}{b_{1}b_{3}}-\frac{a^{2}}{b_{1}^{2}}=-p+3 \zeta
H-(\omega^{de}+\gamma)\rho^{de}  \label{eq6} \\
& \frac{\ddot{b_{1}}}{b_{1}}+\frac{\ddot{b_{2}}}{b_{2}}+\frac{\dot{b_{1}}%
\dot{b_{2}}}{b_{1}b_{2}}-\frac{a^{2}}{b_{1}^{2}}=-p+3 \zeta
H-(\omega^{de}+\eta)\rho^{de}  \label{eq7} \\
& \frac{\dot{b_{1}}\dot{b_{2}}}{b_{1}b_{2}}+\frac{\dot{b_{2}}\dot{b_{3}}}{%
b_{2}b_{3}}+\frac{\dot{b_{3}}\dot{b_{1}}}{b_{3}b_{1}}-\frac{3a^{2}}{b_{1}^{2}%
}=\rho+\rho^{de}  \label{eq8} \\
& 2\dfrac{\dot{b_{1}}}{b_{1}}-\dfrac{\dot{b_{2}}}{b_{2}}-\dfrac{\dot{b_{3}}}{%
b_{3}} =0  \label{eq9}
\end{align}
where an over dot over the field variable represents the derivatives with respect to the cosmic time $t$. Moreover, the unit of cosmic time is considered as, 1 unit of cosmic time = 10 billion years. The average scale factor $R$ and volumetric scale factor $V$ for the model are respectively $R=(b_1b_2b_3)^\frac{1}{3}$ and $V=R^3=b_1b_2b_3$. The generalized mean Hubble parameter $H$ can be expressed as, $H=\frac{\dot{R}}{R}=\frac{1}{3}(H_x+H_y+H_z)$, where $H_x=\frac{\dot{b_1}}{b_1}, H_y=\frac{\dot{b_2}}{b_2}$ and $H_z=\frac{\dot{b_3}}{b_3}$ are the directional Hubble parameters in the direction of $x$, $y$ and $z$ respectively. Now, the field equations \eqref{eq5}-\eqref{eq9} can be framed in Hubble terms as,

\begin{align}
& \dot{H_{y}}+ \dot{H_{z}}+ H_{y}^{2} + H_{z}^{2} + H_{y}H_{z}-\frac{%
a^{2}}{b_1^{2}} = -\bar{p} - (\omega^{de}+ \delta)\rho^{de}  \label{eq10} \\
& \dot{H_{x}}+ \dot{H_{z}}+ H_{x}^{2} + H_{z}^{2} + H_{x}H_{z}-\frac{%
a^{2}}{b_1^{2}} = -\bar{p} - (\omega^{de}+ \gamma)\rho^{de}  \label{eq11} \\
& \dot{H_{x}}+ \dot{H_{y}}+ H_{x}^{2} + H_{y}^{2} + H_{x}H_{y}-\frac{%
a^{2}}{b_1^{2}} = -\bar{p} - (\omega^{de}+ \eta)\rho^{de}  \label{eq12} \\
& H_{x}H_{y}+ H_{y}H_{z}+ H_{z}H_{x}-\frac{3 a^{2}}{b_1^{2}} = \rho +
\rho^{de}  \label{eq13} \\
& 2H_{x}-H_{y}-H_{z}=0,  \label{eq14}
\end{align}

The energy conservation equation for viscous fluid, $T^{(vis)ij}_{;j}=0$ ,
yields

\begin{equation}  \label{eq15}
\dot{\rho}+3(\bar{p}+\rho)H=0
\end{equation}
The energy conservation equation for dark energy fluid, $T^{ij(de)}_{;j}=0$,
yields

\begin{equation}  \label{eq16}
\dot{\rho}^{de}+3 \rho^{de}(\omega^{de}+1)H+ \rho^{de}(\delta H_x+\gamma
H_y+\eta H_z)=0
\end{equation}
From \eqref{eq15}, incorporating the relation between Hubble parameter and average scale factor we get,   
\begin{align}  \label{eq17}
\rho=\rho_{0} R^{-3(\epsilon +1)}= \rho_{0} (b_{1} b_{2} b_{3})^{-(\epsilon
+1)},
\end{align}
where $\rho_0$ is the integration constant or rest energy density.\newline

From \eqref{eq13}, we have, 
\begin{equation}  \label{eq18}
\rho^{de}= H_{x}H_{y}+H_{y}H_{z}+H_{z}H_{x}-\dfrac{3 a^{2}}{b_{1}^2}- \rho
\end{equation}
In order to solve eqn. \eqref{eq16}, we split the conservation equation into two parts one
corresponds to the deviation of equation of the state parameters as $\rho^{de}(\delta H_x+\gamma H_y+\eta H_z)=0$ and other one is the deviation free part as  $\dot{\rho}^{de}+3 \rho^{de}(\omega^{de}+1)H=0$ \cite{akarsu3}. It can be observed that the behaviour of energy density $\rho^{de}$ is controlled by the deviation free part of EoS parameter whereas anisotropic pressure along different directions can be obtained from second part of the above conservation equation as it corresponds to the conservation of matter field with equal pressure along all directions. Hence, we obtained the dark energy density as,
\begin{equation}  \label{eq19}
\dot{\rho}^{de}+3\rho^{de}(\omega^{de}+1)H=0 \Rightarrow
\rho^{de}=\rho_{0}^{de} R^{-3(\omega^{de}+1)}
\end{equation}

Now from \eqref{eq12}, incorporating the value of $\eta$, we get,

\begin{align}  \label{eq20}
-\omega^{de} \rho^{de}= & \biggl(\frac{H_y+H_z}{3H}\biggr)\frac{\ddot{b_{1}}}{%
b_{1}}+\biggl(\frac{H_x+H_z}{3H}\biggr)\frac{\ddot{b_{2}}}{b_{2}}+\biggl(%
\frac{H_x+H_y}{3H}\biggr)\frac{\ddot{b_{3}}}{b_{3}}+ \\
& \frac{H_z}{3H}\frac{\dot{b_{1}}\dot{b_{2}}}{b_{1}b_{2}}+\frac{H_x}{3H}%
\frac{\dot{b_{2}}\dot{b_{3}}}{b_{2}b_{3}}+\frac{H_y}{3H}\frac{\dot{b_{1}}%
\dot{b_{3}}}{b_{1}b_{3}}-\frac{a^{2}}{b_{1}^{2}}+p-3 \zeta H  \notag
\end{align}

Again from eqn. \eqref{eq9}, with the choice of integrating constant to be unity, we get, $b_{1}^2=b_{2}b_{3}$. Moreover for an anisotropic relation, we assume $b_{2}=b_{3}^m$, where $m$ is the average anisotropy parameter \cite{mishra2}. Hence,
\begin{equation}  \label{eq21}
b_{1}=b_{3}^{\frac{m+1}{2}}
\end{equation}

Now, the dark energy density and effective EoS parameter with the function $\Phi(H)= \left(\dfrac{2 \dot{H}+3H^{2}}{m+1}\right)$ can be reformulated respectively as 

\begin{align}  \label{eq22}
\rho^{de}=\left[ 2 \dfrac{(m^{2}+4m+1)}{(m+1)^{2}} \right] H^{2}- \dfrac{3
a^{2}}{b_{1}^{2}}- \rho
\end{align}

\begin{align}  \label{eq23}
\omega^{de} \rho^{de}= -\dfrac{2}{3} \left( \dfrac{m^{2}+4m+1}{m+1} \right)
\Phi(H) + \dfrac{a^2}{b_{1}^{2}}-\bar{p},
\end{align}

With the help of eqns. \eqref{eq17}-\eqref{eq20}, eqns. \eqref{eq10}-\eqref{eq12} can be expressed in functional form as: 
\begin{eqnarray}  \label{eq25}
\gamma = \left(\frac{5+m}{6 \rho^{de}} \right) \chi(m) F(H) \\
\eta = - \left( \frac{5m+1}{6 \rho^{de}} \right) \chi(m) F(H) \\
\delta = - \left( \frac{m-1}{3 \rho^{de}} \right) \chi(m) F(H),  \label{eq26}
\end{eqnarray}

where, $\chi(m)= \dfrac{m-1}{m+1},$ $F(H)= \Phi(H)+\dfrac{3H^{2}}{m+1}$ .

\section{Cosmological models and its behaviour}
From the above formalism it is quite clear that obtaining an exact solution to the field equations is a  cumbersome process. Therefore without violating any physical meaning of the expression and in order to study the cosmological model in this formalism, we have assumed two scale factors one leads to power law expansion and the other to de Sitter expansion.

\subsection{Power law expansion model}
Recently many observational results as well as experiments predict a tensor-to-scalar ratio that provides a convincing results for standard inflationary scenario even though it contradicts the limits from Planck
data. During a power law expansion, the inflationary scenario predicts the generation of gravitational waves. In this model, the scale factor for power law cosmology can be represented as $R=t^k$, where $k$ is a positive constant and $k={\left(\frac{m+1}{2} \right) n}$. Also, $m$ and $n$ are positive constants. Now, the volume scale factor,$V = t^{3k}$ and Hubble parameter $H =\frac{k}{t}$. It is obvious that, for $k > 1,$ the model will be an accelerating one. Now, subsequently the directional Hubble parameters can be obtained as $H_{x} = \left(\frac{m+1}{2} \right)\frac{n}{t}$, $ H_{y} = \frac{mn}{t}$ and $ H_{z} =\frac{n}{t}$ and consequently, the mean deceleration parameter becomes $q = -1 + \frac{2}{n(m+1)}$. The deceleration parameter is a negative constant quantity for $n > \frac{2}{m+1}$, since $m$ and $n$ are positive constants and is good agreement with the present observational data that predicts an accelerating universe, therefore in order to get an accelerating model with this power law scale factor, the exponent $n>1$ if $m < 1$ otherwise it has to be decided from $n>\frac{2}{m+1}$. \\

The universe in general is isotropic, but recent observations from CMB temperature anisotropy, the anisotropic nature is favoured. However, any anisotropy in spatial expansion must be considered as a little perturbation of the isotropic behaviour which suggests that the exponent $m$ must be close to 1. In fact, according to the present result from the analysis of anisotropy as predicted from Planck data \cite{J,K} and from our earlier work , $m\approx 1.0001633$ \cite{mishra2, mishra3}. The power law model is quite successful in the sense that it neither encounter the horizon problem nor witness the flatness problem with $n \geqslant \frac{2}{m+1}$. The energy density contribution coming from the usual cosmic fluid for the power law model reduces to,

\begin{align}  \label{eq27}
\rho= \frac{\rho_{0}}{t^{\frac{3}{2}(m+1)(\epsilon+1)n}}
\end{align}

Now, with the help of eqn. \eqref{eq27}, the dark energy density and dark energy EoS parameter as described in \eqref{eq22} and \eqref{eq23} can be respectively reduced to,

\begin{align*}  
\rho^{de}= \left[ \dfrac{(m^{2}+4m+1)n^{2}}{2} \right] \dfrac{1}{t^{2}} - 
\dfrac{3 a^{2}}{t^{n(m+1)}}-\dfrac{\rho_{0}}{t^{\frac{3}{2}%
(m+1)(\epsilon+1)n}}
\end{align*}
and 
\begin{equation}\label{eq28}
\omega ^{de}=\dfrac{1}{\rho ^{de}}\left[ \left\{ \dfrac{%
n(m^{2}+4m+1)(4-3n(m+1))}{6(m+1)}\right\} \dfrac{1}{t^{2}}+\dfrac{a ^{2}%
}{t^{n(m+1)}}-\bar{p}\right] ,
\end{equation}

where, $\bar{p}=\epsilon \rho _{0}^{de}$. The skewness parameters $\delta ,$ 
$\gamma $ and $\eta $ reduce to 
\begin{eqnarray} 
\gamma  &=&\frac{\left( 5+m\right) \left( m-1\right) }{3\left( m+1\right)
^{2}}\left( \frac{3k^{2}-k}{t^{2}}\right) \dfrac{1}{\rho ^{de}}\label{eq29} \\
\eta  &=&-\frac{\left( 5m+1\right) \left( m-1\right) }{3\left( m+1\right)
^{2}}\left( \frac{3k^{2}-k}{t^{2}}\right) \dfrac{1}{\rho ^{de}} \label{eq30}\\
\delta  &=&-\frac{2\left( m-1\right) ^{2}}{3\left( m+1\right) ^{2}}\left( 
\frac{3k^{2}-k}{t^{2}}\right) \dfrac{1}{\rho ^{de}}\label{eq31}
\end{eqnarray}

where $k=\frac{n(m+1)}{2}$. It is seen that both $\rho$ and $\rho^{de}$ decrease with the increase in time. The decrease in $\rho^{de}$ is decided by three different factors i.e $t^{-2}$ in the first term , $t^{-n(m+1)}$ in the second term and $t^{3/2 (1+ \epsilon)(m+1)n}$ in the third term. The role of bulk viscous cosmic fluid comes through the third term. One may note that, if $\epsilon =-1$, even though the contribution coming from the usual cosmic fluid does not vanish, it does not contribute to the time variation of the dark energy density. For $\epsilon = -\frac{1}{3}$, the time variation of second and third terms can be clubbed together. Consequently, for this choice the skewness parameters becomes a constant quantity appears to be a simple time independent deviations from usual isotropic pressure.  

\begin{figure}[tbph]
\minipage{0.32\textwidth}
\centering
\includegraphics[width=\textwidth]{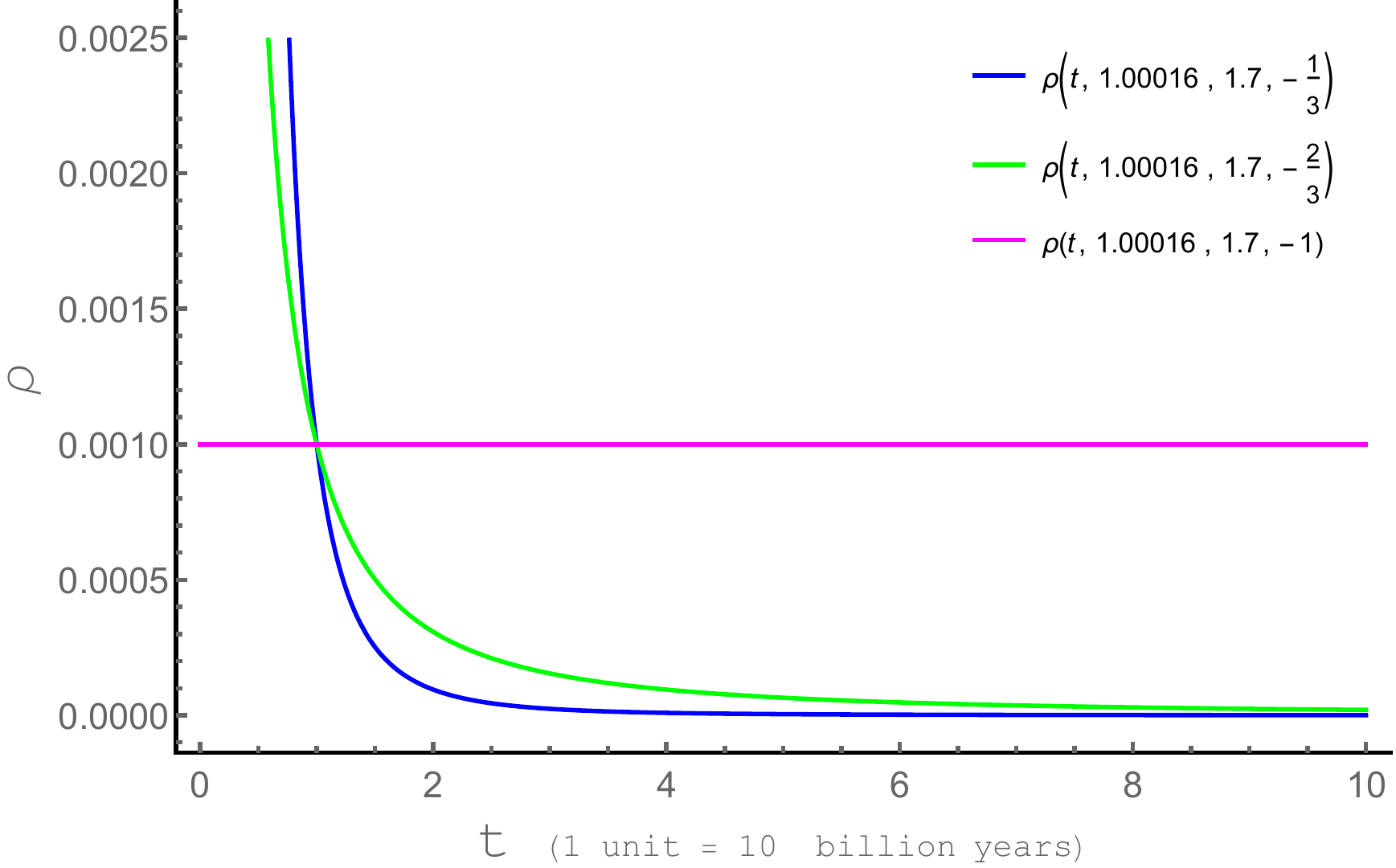} 
\caption{ Variation of $\rho$ versus  $t$  for representative values of the parameter $m=1.0001633, n=1.7, \rho_0=0.001$ with different $\epsilon=-\frac{1}{3},-\frac{2}{3}, -1$}
\endminipage
\minipage{0.32\textwidth}
\centering
\includegraphics[width=\textwidth]{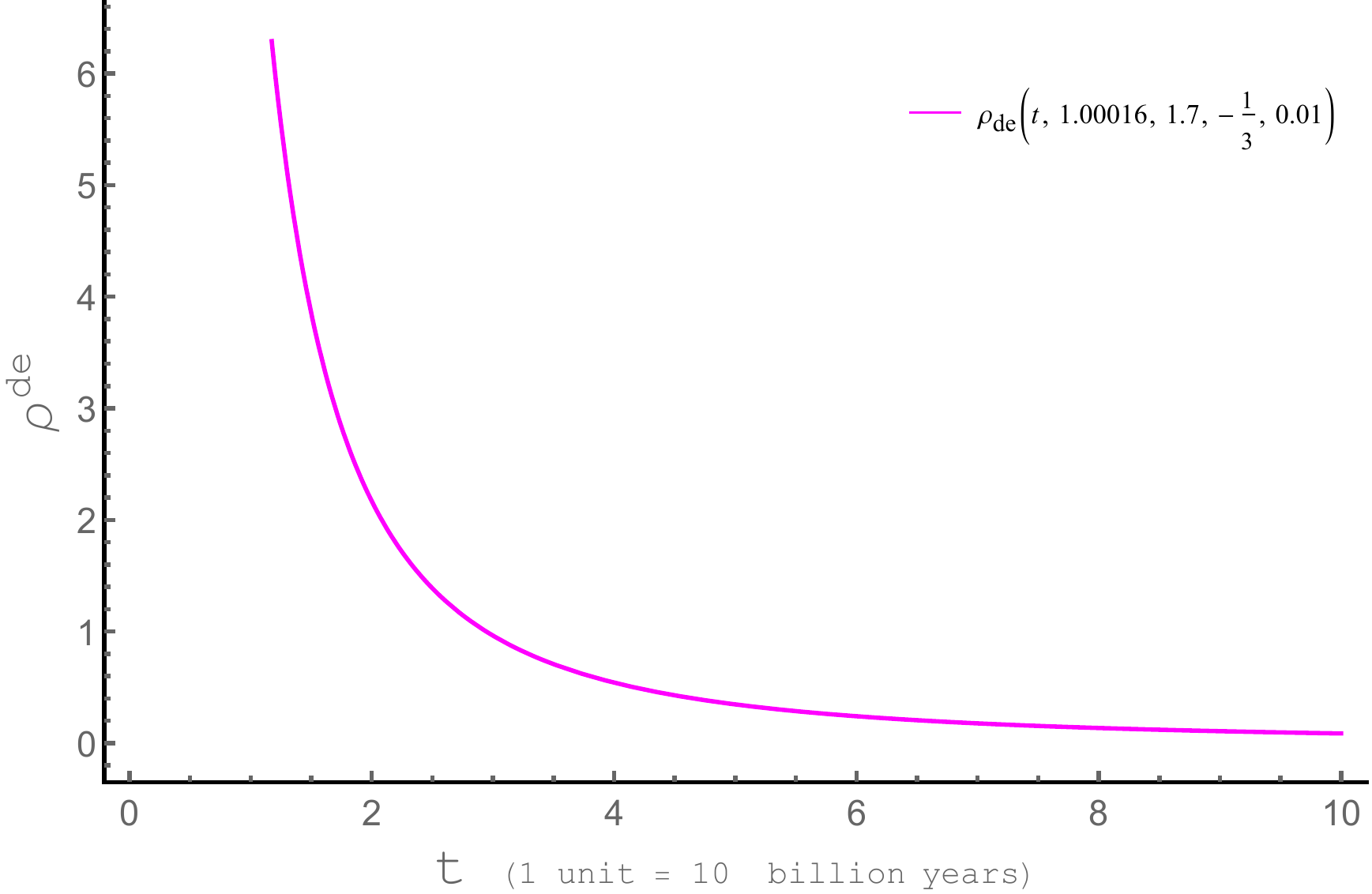}
\caption{Variation of $\rho^{de}$ versus  $t$ for representative values of the parameter $m=1.0001633, n=1.7, a=0.01, \rho_0=0.001$ with different $\epsilon=-\frac{1}{3}$
}
\endminipage
\minipage{0.32\textwidth}
\centering
\includegraphics[width=\textwidth]{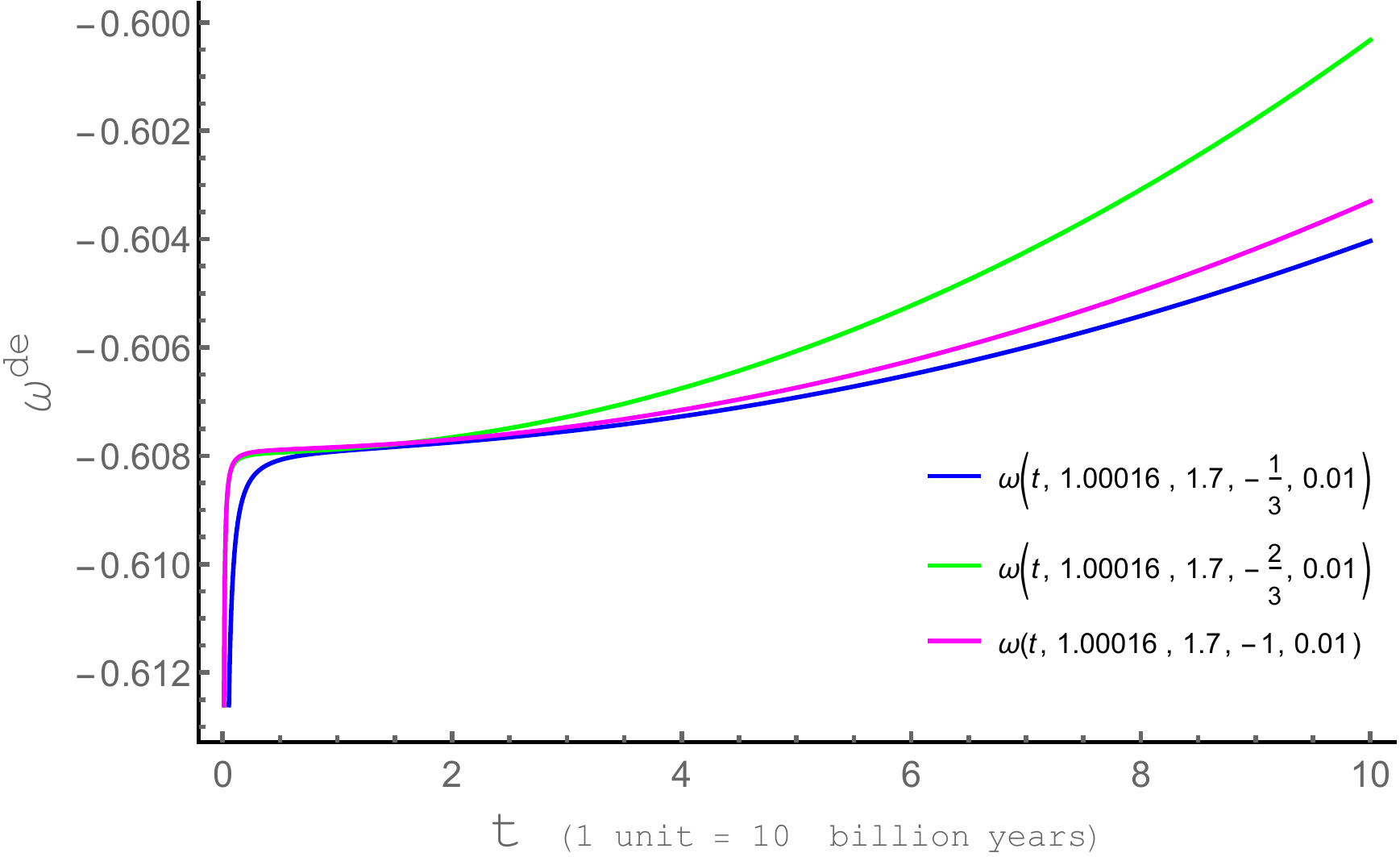} 
\caption{Variation of $\omega^{de}$ versus  $t$ for representative values of the parameter $m=1.0001633, n=1.7, a=0.01, \rho_0=0.001$ with different $\epsilon=-\frac{1}{3},-\frac{2}{3}, -1$
}
\endminipage
\end{figure}
The figures in the manuscript have been drawn for different Physical quantities which are expressed in Planckian unit system $(c=G=k_B=h=1)$. Also, 1 unit of cosmic time = 10 billion years. In FIG. 1 and FIG. 2, we have observed respectively that the matter energy density $\rho$ and the dark energy density $\rho^{de}$ remains positive during the cosmic evolution for the representative value of the constants ($m=1.0001633,n=1.7,a=0.01, \rho_0=0.001$). Hence, it indicates that both weak energy condition (WEC) and null energy condition (NEC) are satisfied in the derived model. Further, both $\rho$ and $\rho^{de}$ decreases with increase in time and slowly reaches to a small positive values in the present epoch. The value of dark energy density comes closer to zero and then smoothly approaches to small positive value which indicates the considered two fluid affects the dark energy density. It is worthy to note here that irrespective of the value of the viscous coefficient $\epsilon$, the behaviour of $\rho^{de}$ remains alike. So, in FIG.2, we have chosen the value of the viscous coefficient to be $-\frac{1}{3}$. However, a small effect of viscous fluid in dark energy density cannot be ruled out. FIG. 3, represents the variation of $\omega^{de}$ with cosmic time for different values of viscous coefficients $\epsilon$. The range value of EoS parameter suggested by combination of SNIa data with CMB anisotropy and galaxy clustering statistics is [-1.33,-0.79] \cite{suresh11}; whereas the range suggested by recent observations are reduced to more stringent constraints around -1  \cite{planck15,H,I}. However, we consider here the earlier data range since power law behaviour dominates the cosmic dynamics in early phase of cosmic evolution \cite{mishra2}. For $\epsilon=-1/3,-2/3,-1$, FIG. 3 clearly shows that $\omega$ evolves with in a range, which is almost aligned with SN Ia and CMB observations. Moreover, it is observed that when the bulk viscous coefficients increases, the EoS parameter gradually converges to $\Lambda_{CDM}$ at late time.\\

\begin{figure}[tbph]
\minipage{0.32\textwidth}
\centering
\includegraphics[width=\textwidth]{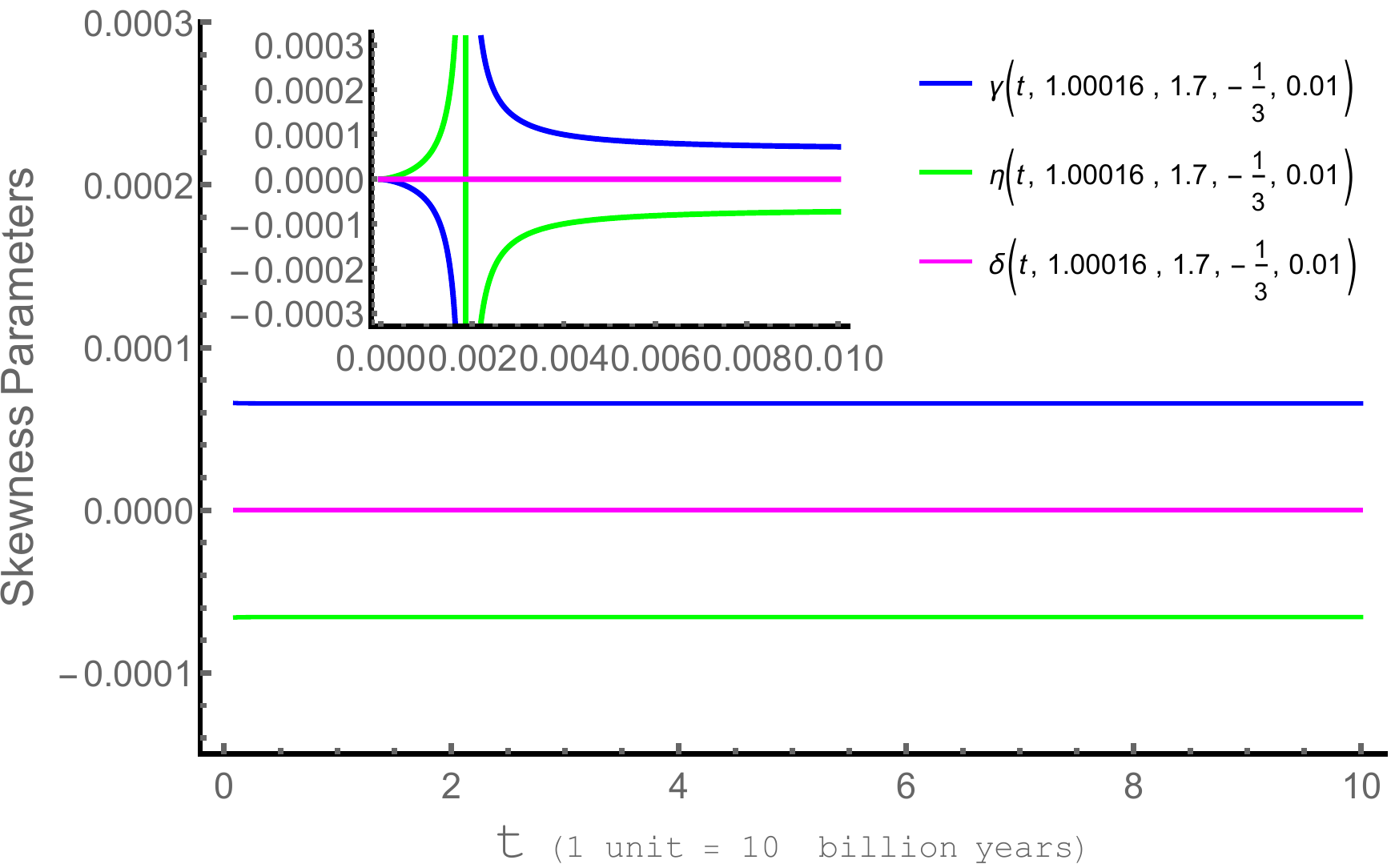}
\caption{Variation of $\delta, \gamma, \eta$ versus $t$ for representative values of the parameter $m=1.0001633, n=1.7, a=0.01, \rho_0=0.001$ for $\epsilon=-\frac{1}{3}$
}
\endminipage
\minipage{0.32\textwidth}
\centering
\includegraphics[width=\textwidth]{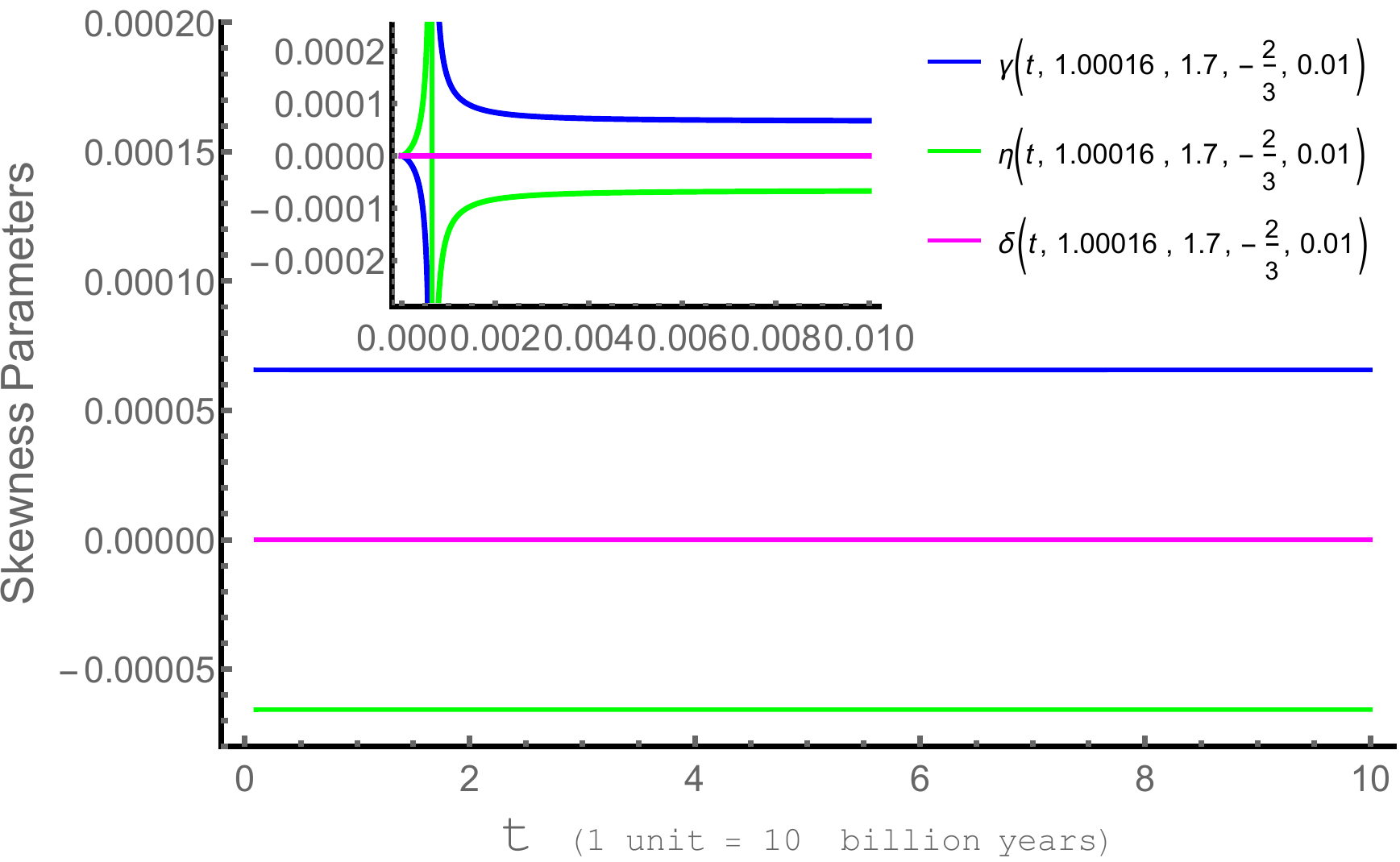} 
\caption{Variation of $\delta, \gamma, \eta$ versus $t$ for representative values of the parameter $m=1.0001633, n=1.7, a=0.01, \rho_0=0.001$ for $\epsilon=-\frac{2}{3}$}
\endminipage
\minipage{0.32\textwidth}
\centering
\includegraphics[width=\textwidth]{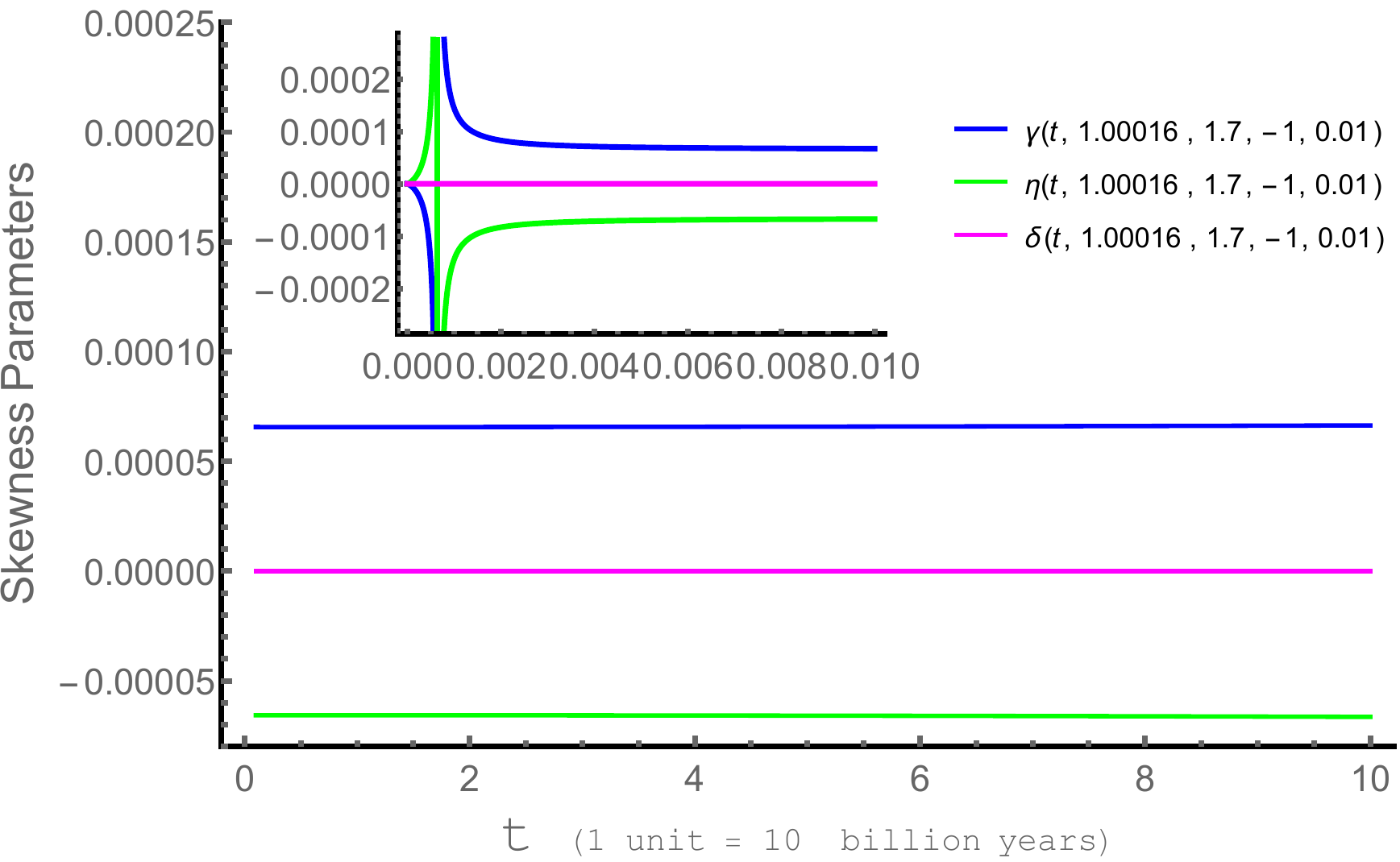}
\caption{Variation of $\delta, \gamma, \eta$ versus $t$ for representative values of the parameter $m=1.0001633, n=1.7, a=0.01, \rho_0=0.001$ for $\epsilon=-1$}
\endminipage
\end{figure}

The behaviour of the skewness parameters obtained in eqns. \eqref{eq29}-\eqref{eq31} has been graphically represented in FIG. 4, FIG. 5 and FIG. 6 respectively for $\epsilon=-1/3$, $\epsilon=-2/3$ and $\epsilon=-1$. In these figures, it can be noted that the behaviour of skewness parameter are totally controlled by the behaviour of the anisotropic parameter $m$ \cite{mishra2, tripa1,ppr1, ppr2}. For $m=1$, the skewness parameters vanish giving an indication that the viscous matter affects the skewness parameter. We have observed that, at an early cosmic phase, $\eta$ starts with a negative value far from zero, increases with the cosmic time to became maximum and then becomes constant with further increase in cosmic time. The skewness parameter $\delta$ also starts from a small negative value close to zero and becomes constant with respect to cosmic time. The evolutionary behaviour of $\gamma$ is just the mirror image of $\eta$. It can be noted that the pressure anisotropy factors along $x$, $y$ and $z$ axis $(\eta, \delta,\gamma)$ evolves with different nature, attain their extreme values in a definite range of cosmic time $0.001<t<0.003$ and remain constant at later time.Therefore, it can be inferred that, in power law cosmology, at an early phase of cosmic evolution, the pressure was assumed to be isotropic; however in the late phase, pressure anisotropies still remain. It can also be noted that, the behaviour of skewness parameters are independent of the choice of the bulk viscous coefficient.

\subsection{de Sitter expansion model}

In de Sitter model, the scale factor is taken as $R=e^{(\frac{m+1}{2})\xi t}$, where $\xi$ is a positive constant. In this model, the Hubble parameter is a constant quantity and remains the same through out the cosmic evolution. The directional Hubble rates along different spatial directions are also constants and can be expressed as $H_{x}=\left(\frac{m+1}{2} \right) \xi,$ $H_{y}=m \xi,$ $H_{z}= \xi$. So, $\xi$ can be expressed as $\xi= \frac{2H}{m+1}$. With this assumption of the scale factor, the energy density contribution coming from the usual cosmic bulk viscous fluid for the de Sitter model reduces to, 
\begin{equation}
\rho =\frac{\rho _{0}}{e^{\frac{3}{2}(1+\epsilon )(m+1)\xi t}}  \label{eq32}
\end{equation}%
The energy density increases with the decrease in the value of $\epsilon $ and vice-versa. For the particular choice $\epsilon =-1,$ $\rho $ becomes independent of time and assumes a constant value $\rho _{0}$ throughout the cosmic evolution (FIG.7).\\
The rest energy density $\rho^{de}$and the dark energy EoS parameter $\omega ^{de}$ for the de Sitter universe can now be obtained as,

\begin{equation}
\rho ^{de}=\xi ^{2}\left( \dfrac{m^{2}+4m+1}{2}\right) -\dfrac{3a^{2}}{%
e^{\xi (m+1)t}}-\dfrac{\rho _{0}}{e^{\frac{3}{2}(1+\epsilon )(m+1)\xi t}}
\label{eq33}
\end{equation}

\begin{equation}
\omega ^{de}=\dfrac{1}{\rho ^{de}}\left[ -\dfrac{\left( m^{2}+4m+1\right)
\xi ^{2}}{2}+\dfrac{a^{2}}{e^{\xi (m+1)t}}-\bar{p}\right]  \label{eq34}
\end{equation}

where, $\bar{p}=\epsilon \rho _{0}^{de}$. Subsequently, the skewness parameters $\delta ,$ 
$\gamma $ and $\eta $ can be expressed as 
\begin{eqnarray}
\gamma  &=&\frac{\left(5+m\right) \left( m-1\right) }{4}\left(\frac{\xi^{2}}{\rho
^{de}}\right) \label{35}\\
\eta  &=&-\frac{\left(5m+1\right) \left( m-1\right)}{4}\left(\frac{\xi^{2}}{\rho
^{de}}\right) \label{36}\\
\delta  &=&-\frac{2\left( m-1\right) ^{2}}{4}\left(\frac{\xi ^{2}}{\rho ^{de}}\right) \label{37}
\end{eqnarray}

In the de Sitter model, the dark energy density decreases with increase in time and asymptotically reduces to a positive constant. The decrement in $\rho^{de}$ is decided by four different  factors in the second and third term viz., $m, \xi, a, \rho_0$. The role of bulk viscous cosmic fluid comes through the third term. The contribution from the bulk viscous cosmic fluid becomes time independent for $\epsilon=-1$. The pressure anisotropies defined earlier in  different axis depend on the behaviour of skewness parameters whereas the dark energy density $\rho_{de}$ depends on the barotropic equation of state $\epsilon$ (FIG. 8).\\

\begin{figure}[tbph]
\minipage{0.32\textwidth}
\centering
\includegraphics[width=\textwidth]{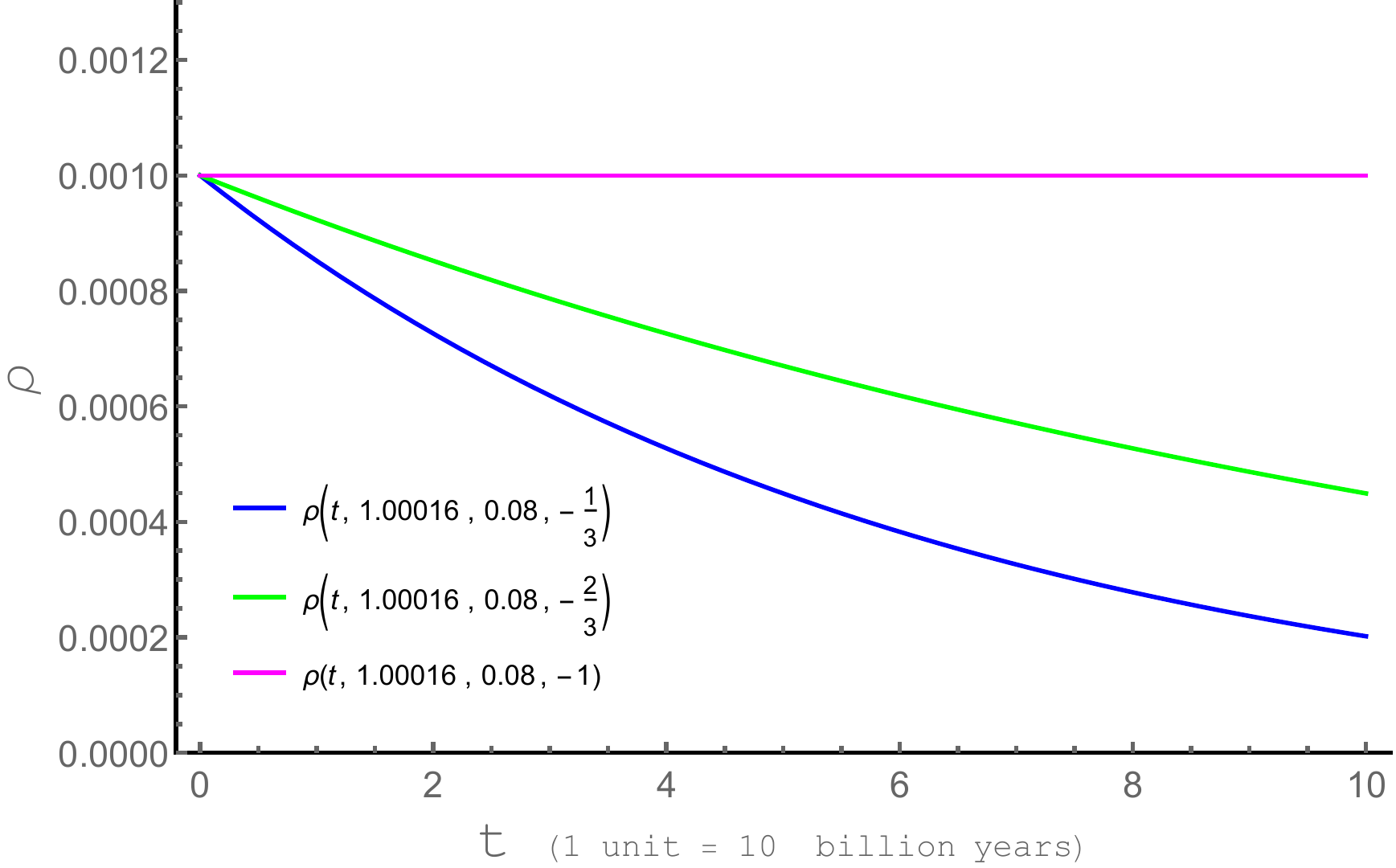} 
\caption{Variation of $\rho$ versus $t$ for representative values of the parameter $m=1.0001633, \rho_0=0.001, \xi=0.08$ with different $\epsilon=-\frac{1}{3},-\frac{2}{3}, -1$}
\endminipage
\minipage{0.32\textwidth}
\centering
\includegraphics[width=\textwidth]{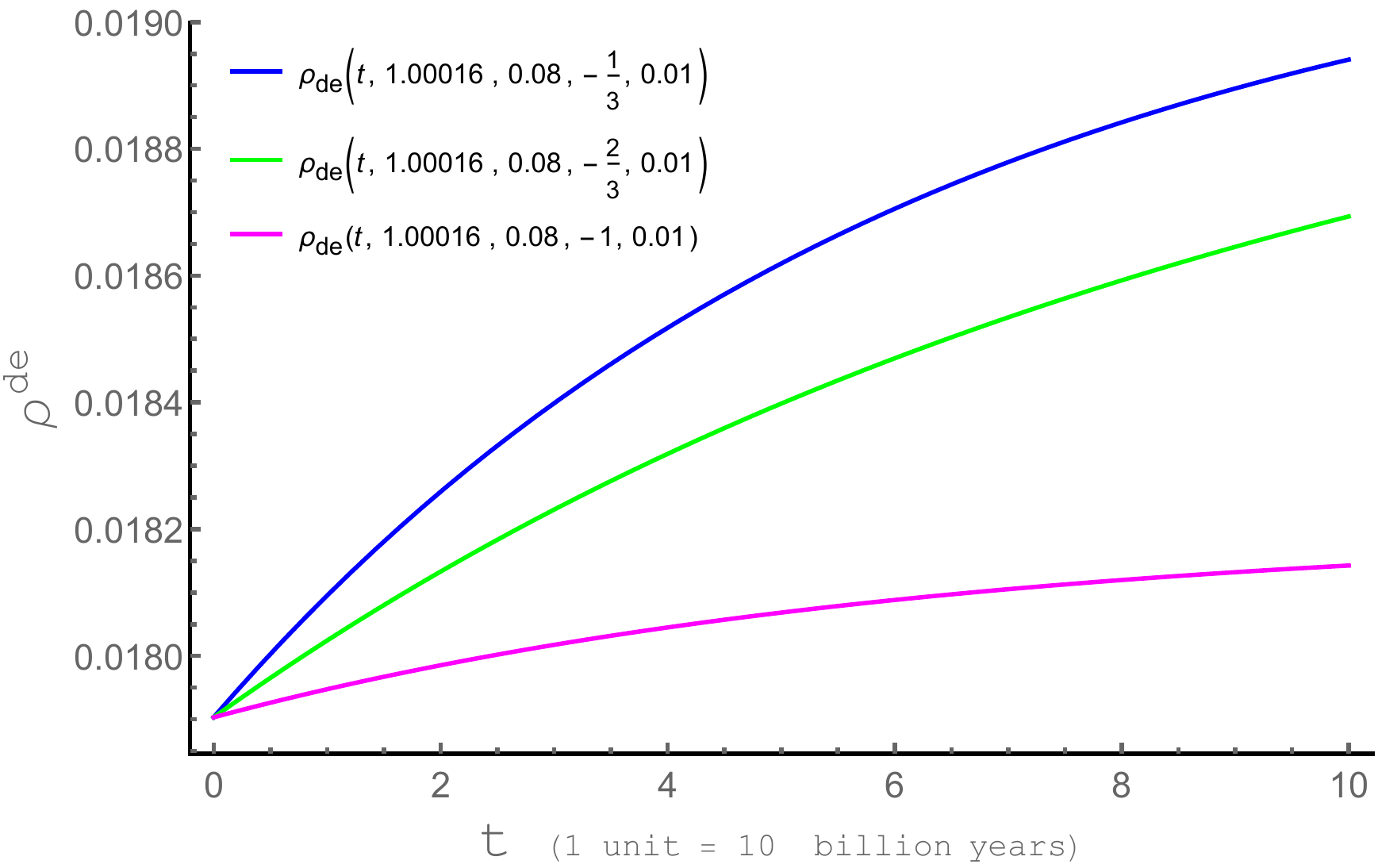}
\caption{Variation of $\rho^{de}$ versus $t$ for representative values of the parameter $m=1.0001633, \rho_0=0.001,a=0.01, \xi=0.08$ with different $\epsilon=-\frac{1}{3},-\frac{2}{3}, -1$}
\endminipage
\minipage{0.32\textwidth}
\centering
\includegraphics[width=\textwidth]{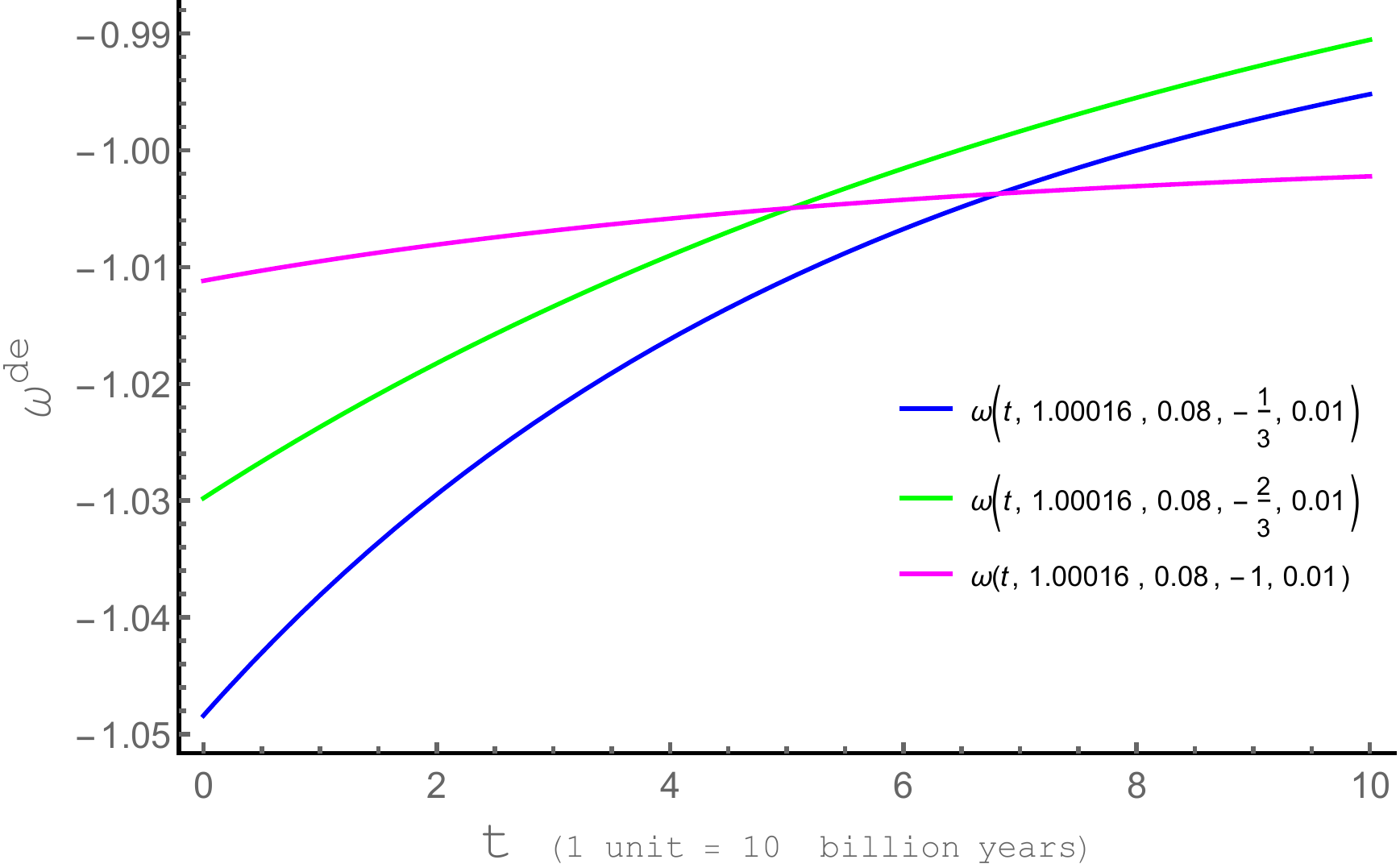} 
\caption{Variation of $\omega^{de}$ versus $t$ for representative values of the parameter $m=1.0001633, \rho_0=0.001,a=0.01, \xi=0.08$ with different $\epsilon=-\frac{1}{3},-\frac{2}{3}, -1$}
\endminipage
\end{figure}
The EoS parameter in FIG. 9 is very much sensitive to the choice of $\epsilon$. For different choices, they starts from different values at the early phase and maintain the same evolutionary state at late phase falling in the observed range as obtained in 2015 Planck data $\omega=-1.019^{+0.075}_{-0.080}$ \cite{planck15}.   The behaviour of EoS parameters are directly proportional to the increasing values of viscous coefficient $\epsilon$ at early phase but behave differently at late phase, gathered some amount of energy at early phase. The reason being the dynamics of EoS is greatly affected at early phases is that the bulk viscous has a substantial contribution to the density parameter at that corresponding phase. But at late phase, the dark energy dominates in spite of the presence of bulk viscous fluid. Hence, cosmic bulk viscous fluid has a very little impact on the dynamics of EoS parameter\cite{brevik}.\\

\begin{figure}[tbph]
\minipage{0.32\textwidth}
\centering
\includegraphics[width=\textwidth]{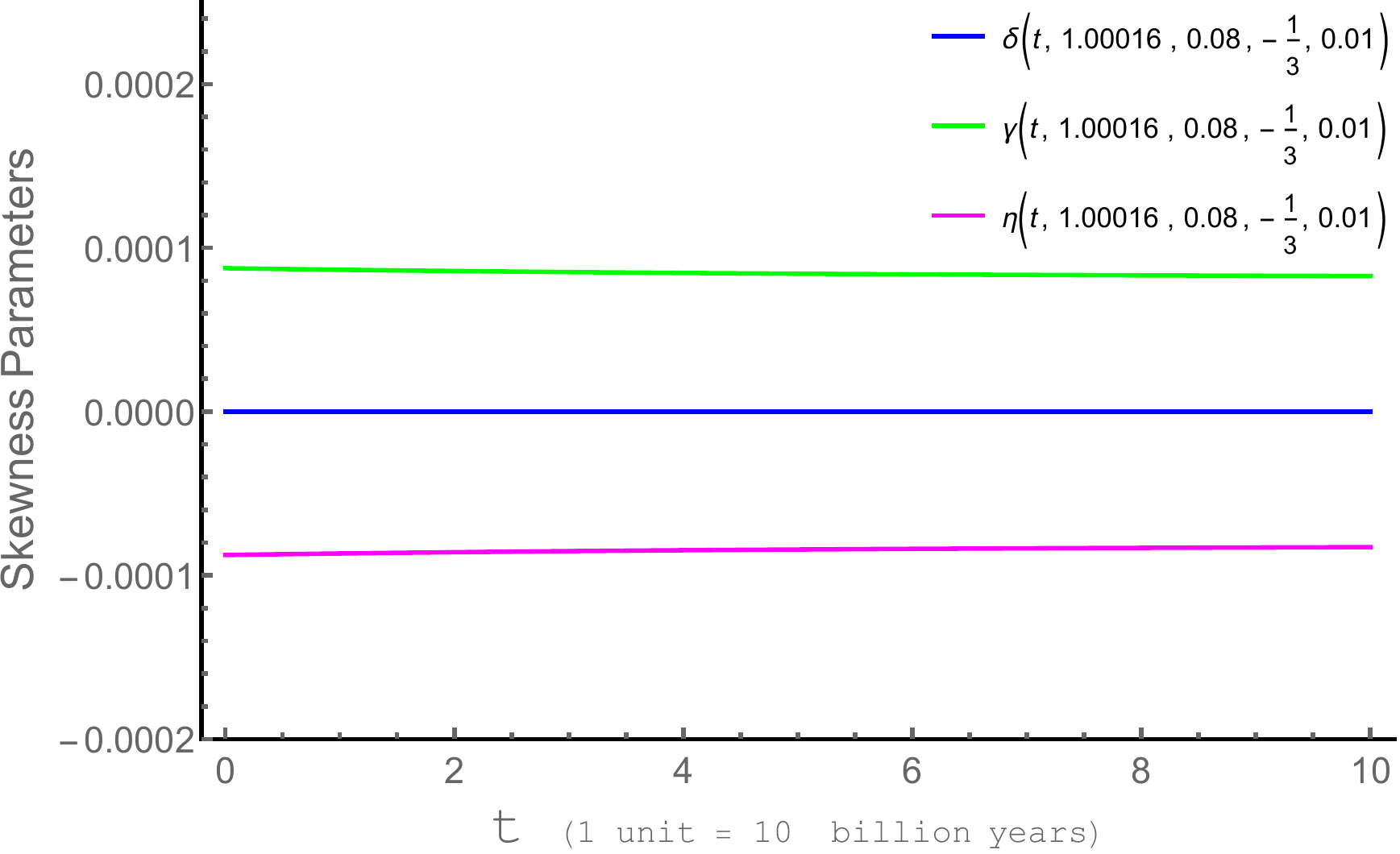}
\caption{Variation of $\delta, \gamma, \eta$ versus $t$ for representative values of the parameter $m=1.0001633,  \rho_0=0.001,\alpha=0.01, \xi=0.08$ for $\epsilon=-\frac{1}{3}$
}
\endminipage
\minipage{0.32\textwidth}
\centering
\includegraphics[width=\textwidth]{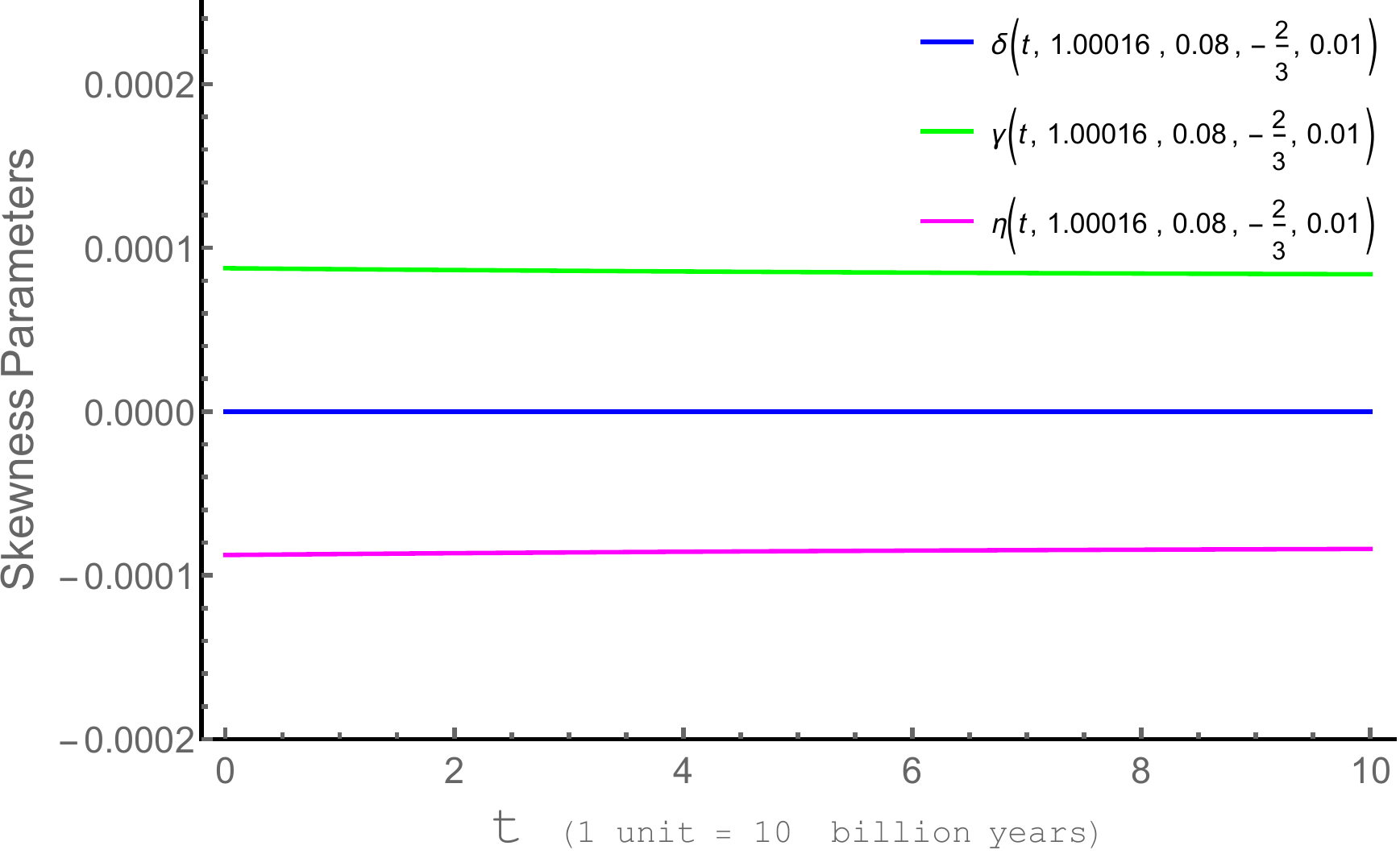} 
\caption{$\delta, \gamma, \eta$ versus $t$ for representative values of the parameter $m=1.0001633, \rho_0=0.001,a=0.01,   \xi=0.08$ for $\epsilon=-\frac{2}{3}$}
\endminipage
\minipage{0.32\textwidth}
\centering
\includegraphics[width=\textwidth]{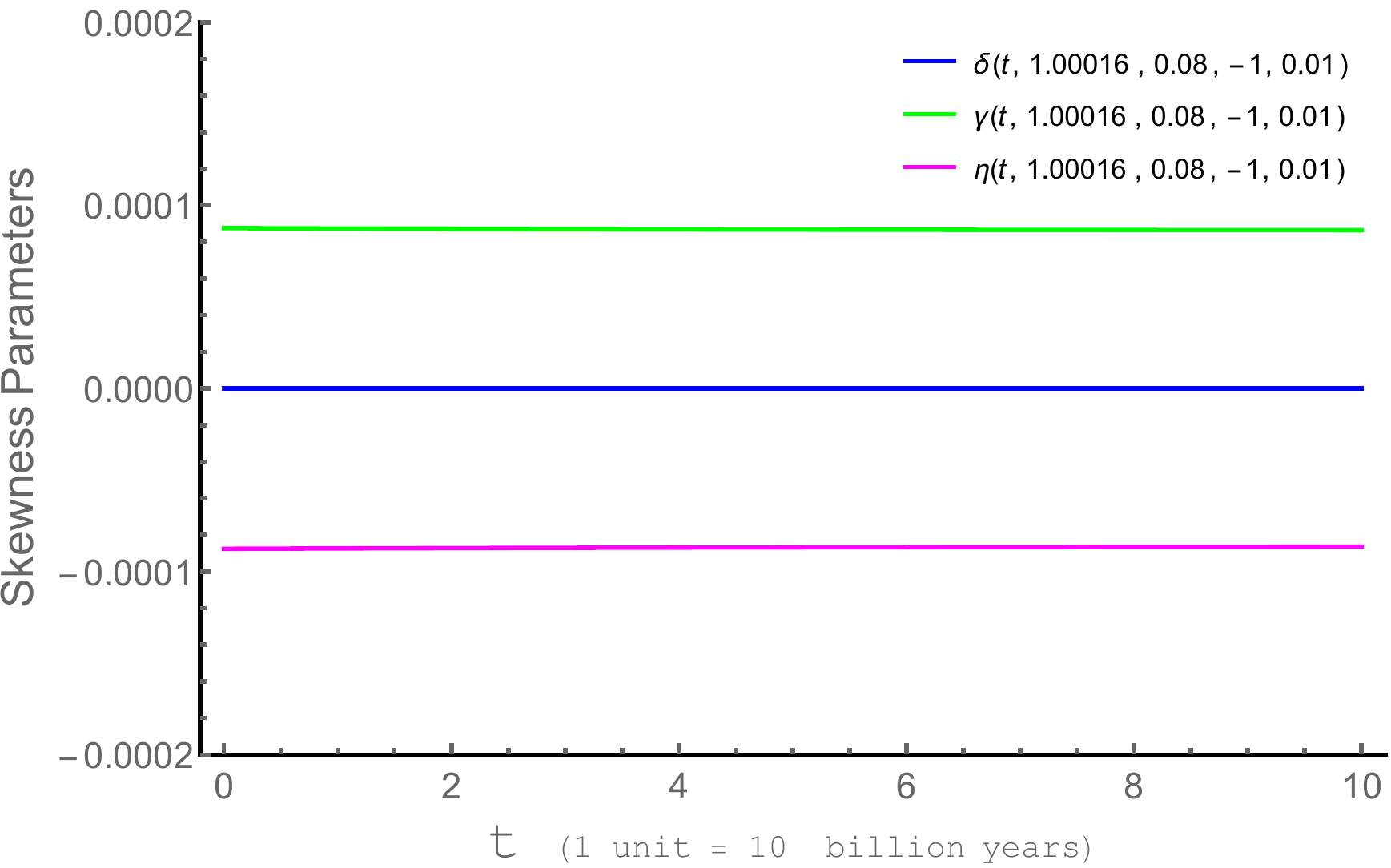}
\caption{$\delta, \gamma, \eta$ versus $t$ for representative values of the parameter $m=1.0001633, \rho_0=0.001,a=0.01, \xi=0.08$ for $\epsilon=-1$}
\endminipage
\end{figure}

The DE skewness parameters are plotted as a function of cosmic time for three representative values of bulk viscous coefficient, $\epsilon= -1, - \frac{2}{3}, -\frac{1}{3}$. The corresponding skewness parameters are shown for these three viscous coefficients respectively in FIG. 10, FIG.11 and FIG. 12. With increase of viscous coefficient of matter, skewness parameters show non evolving behaviour in past epoch and rapidly evolve at late phase. The anisotropy in the DE pressure along $x$-direction is almost unaffected by cosmic expansion for all three considered viscous coefficient
values. So, the pressure anisotropy vanishes along $x$- axis. $\delta$'s are less affected by the presence of cosmic fluid compared to $\gamma$'s and $\eta$'s. The DE pressure along $y$ and $z$-directions are mostly affected. The reason behind the sensitivity may be due to the consideration of assuming mean Hubble parameter same as directional Hubble parameter along $x$-axis. At early times, the universe is predicted to have almost isotropic fluid which became anisotropic with the growth of cosmic time. Due to presence of bulk viscous fluid, the anisotropy in DE pressure continues along with the cosmic expansion and decreases slowly at the later period as shown in FIG. 10, FIG.11 and FIG 12. 

\section{Conclusion}

In the present work, we have investigated the role of anisotropic components on the dynamical aspects of DE model in Bianchi V space-time in a two fluid situations. Two cosmological  models have been constructed one pertaining to power law cosmology and the other one to de Sitter universe. The present model favours a quintessence energy dominated universe in the later universe as $-1<\omega^{de}<0$; however in early universe it favours phantom region. In power law model, EoS parameter lies within the predicted range by observational data. In the de Sitter model, DE dominates at late phase of EoS parameter and bulk viscous fluid plays an important role at early universe. The skewness parameters are dynamically evolving with respect to cosmic expansion. In power law, skewness parameter evolves with different values at early phase, where as, it remains constant at late phase, indicating constant anisotropy rate. However, in de Sitter model, the skewness parameters decreases at the later cosmic period and shows a small amount of anisotropy in future cosmic time. The behaviour of skewness parameters are independent of the choice of the bulk viscous coefficient. There is a lot of observational evidences to support the $\Lambda_{CDM}$, such as CMB and redshift-distance relation. However, our considered model is a generalization of FRW model or $\Lambda_{CDM}$ model as discussed in section II. Also, our model is scale factor dependent and may change its behaviour in different scale factors; however the formalism developed here clearly indicates the accelerating behaviour of the expanding universe. Moreover, the resemblance of data of considered model with standard $\Lambda_{CDM}$ model, our model found here also aligned with the present day observational outcomes. \\
  
There are ways to know whether the model approaching to  $\Lambda_{CDM}$ model is accurate or not. One way is $\Lambda_{CDM}$ universes may converge to de Sitter universe under special conditions as de Sitter model has the exponential growth. In our investigation also, the EoS parameter in de Sitter universe is similar to EoS parameter of $\Lambda_{CDM}$ model at late cosmic time of evolution. Another way is the state finder diagnostics, which checks the validity of the model. $\Lambda_{CDM}$ model has state finder pair ${(r,s)}={(1,0)}$. The model discussed here also confirms the acceptability by state finder analysis with the pair ${(1,0)}$ for both the de Sitter universe and for power law for large value of the exponent $m$. 

\section{Competing Interests}
The authors declare that they have no competing interests.

\section*{Acknowledgement}
BM and PPR acknowledge DST, New Delhi, India for providing facilities through DST-FIST lab, Department of Mathematics, where a part of this work was done. SKJP thanks NBHM, Department of Atomic Energy (DAE), Government of India for the post-doctoral fellowship. The authors are thankful to the anonymous referees for their valuable comments and suggestions for the improvement of the paper.


\begin{thebibliography}{999}


\section*{References}
\bibitem{riess} A. G. Riess,"Observational evidence from Supernovae for an accelerating universe and a cosmological constant", \textit{Astrono.J.}, \textbf{116}(3), 1009-1038, (1998).

\bibitem{perm} S. Perlmutter, "Measurements of $\Omega$ and $\Lambda$ from 42 high-redshift supernovae", \textit{\ Astrophy. J.}, \textbf{517} (2), 565-586, (1999).

\bibitem{A} P. A. R. Ade et al., "Planck Collaboration XV, Planck 2013 results. XV. CMB power spectra and
likelihood", \textit{Astronomy Astrophys}, \textbf{571}, A15, (2014).

\bibitem{B} P. A. R. Ade et al., "Planck Collaboration XI, Planck 2015 results. XI. CMB power spectra,
likelihoods  and robustness of parameters", \textit{Astronomy Astrophys}, \textbf{594}, A11,  (2016).

\bibitem{C} Pogosian, L. , Vachaspati, T., "Cosmic microwave background anisotropy from wiggly strings", \textit{Phys. Rev.}, \textbf{D60}, 083504, (1999).

\bibitem{D} P. A. R. Ade et al., "Planck Collaboration XVII, Planck 2013 results. XVII. Gravitational lensing by large-scale structure", \textit{Astronomy Astrophys}, \textbf{571}, A17, (2014). 

\bibitem{E} P. A. R. Ade et al., "Planck Collaboration XV, Planck 2015 results. XV. Gravitational lensing", \textit{Astronomy Astrophys},\textbf{594}, A15, (2016).

\bibitem{cope} E.J. Copeland, M. Sami, S. Tsujikawa, "Dynamics of dark energy", \textit{Int. J. Mod. Phys. D}, \textbf{15}, 1753-1935, (2006).

\bibitem{li} M. Li et al., "Dark energy", \textit{Commun. Theor. Phys.}, \textbf{56}, 525-604, (2011).

\bibitem{kumar1} S. Kumar, A.K. Yadav, "Some Bianchi type-V models of accelerating universe with dark energy", \textit{ Mod. Phys. Lett. A}, \textbf{26}, 647-659, (2011).

\bibitem{barrow1} J.D. Barrow, "Cosmological limits on slightly skew stresses", \textit{Phys. Rev. D}, \textbf{55}, 7451-7460, (1997).

\bibitem{barrow2} J.D. Barrow, R. Maartens,"Anisotropic stresses in inhomogeneous universes", \textit{Phys. Rev. D}, \textbf{59}, 043502, (1999).

\bibitem{F} T. Abbott et. al., "The Dark Energy Survey: more than dark energy – an overview", \textit{Mon. Not. of the Royal Astron. Soc.}, \textbf{460}, 1270, (2016).

\bibitem{G} P. A. R. Ade et al., "Planck Collaboration XIV, Planck 2015 results. XIV. Dark energy and modified gravity", \textit{Astronomy Astrophys},\textbf{594}, A14 (2016).

\bibitem{cruz} M. Cruz et al., "The Non-Gaussian cold spot in the 3 Year Wilkinson Microwave Anisotropy Probe data", \textit{Astrophys. J.}, \textbf{655}, 11-20, (2007).

\bibitem{hoftu} J. Hoftuft et al., "Increasing evidence for hemispherical power asymmetry in the five-year WMAP data", \textit{Astrophys. J.}, \textbf{699}, 985-989, (2009).

\bibitem{bennet1} C.L. Bennett et al, "Seven-year Wilkinson Microwave Anisotropy Probe (WMAP) Observations: Are There Cosmic Microwave Background Anomalies?", \textit{Astrophys. J. Suppl. Ser.}, \textbf{192}, 17, (2011).

\bibitem{spergel} D.N. Spergel et al., "Three-Year Wilkinson Microwave Anisotropy Probe (WMAP) Observations: Implications for Cosmology", \textit{Astrophys. J. Suppl. Ser.}, \textbf{170}, 377-408, (2007).

\bibitem{komat1} E. Komatsu et al., "Seven-year Wilkinson Microwave Anisotropy Probe (WMAP) Observations: Cosmological Interpretation", \textit{Astrophys. J. Suppl. Ser.} \textbf{192}, 18, (2011)

\bibitem{campa4} L. Campanelli et al., "Testing the Isotropy of the Universe with Type Ia Supernovae",\textit{Phys. Rev. D}, \textbf{83}, 103503, (2011).

\bibitem{mishra1} B. Mishra, P. K. Sahoo and S. K. Tripathy, "Pressure anisotropy and dark energy models in scale invariant theory of gravitation", \textit{Astrophysics Space Science}, \textbf{356}, 163-171, (2015).

\bibitem{mishra17} B. Mishra, Pratik P. Ray, SKJ Pacif, "Dark energy cosmological models with general forms of scale factor", \textit{Euro. J Phus Plus}, \textbf{132}, 429, (2017).

\bibitem{tripa1} S.K. Tripathy, B.Mishra, P.K. Sahoo, "Two fluid anisotropic dark energy
models in a scale invariant theory", \textit{Euro. J Phus Plus}, \textbf{132}, 388, (2017).

\bibitem{ppr1} B. Mishra, P.K. Sahoo, Pratik P. Ray, "Accelerating dark energy cosmological model in two fluid with hybrid scale factor", \textit{Int J. of Geom. Methods in Mod. Phys}, \textbf{14},1750124, (2017).

\bibitem{ppr2} B. Mishra, S. K. Tripathy, Pratik P. Ray, "Anisotropy in Dark Energy",: \textit{arXiv:1701.08632v2 [physics.gen-ph]}, (2017).

\bibitem{akarsu3} O. Akarsu, C. B. Kilinc, "LRS Bianchi type I models with anisotropic dark energy and constant deceleration parameter", \textit{Gen. Rel. Grav.}, \textbf{42}, 119, (2010).

\bibitem{akarsu4} O. Akarsu, C. B. Kilinc, "Bianchi type III models with anisotropic dark energy", \textit{Gen. Rel. Grav.}, \textbf{42}, 763, (2010).

\bibitem{yadav1} A. K. Yadav, F. Rahaman, S. Ray, "Dark Energy Models with Variable Equation of State Parameter", \textit{Int. J. Theo. Phys.}, \textbf{50}, 871, (2011).

\bibitem{amir1} H. Amirhashchi, "Phantom instability of viscous dark energy in anisotropic space-time", \textit{Astrophys. Space Sci.}, \textbf{345}, 439, (2013).

\bibitem{pradhan1} A. Pradhan, H. Amirhashchi, B. Saha, "An interacting and non-interacting two-fluid scenario for dark energy in FRW universe with constant deceleration parameter" \textit{Astrophys. Space Sci.}, \textbf{333}, 343, (2011).

\bibitem{shey} A. Sheykhi, M. R. Setare, "Thermodynamical Description of the Interacting new agegraphic dark energy", \textit{Mod. Phys. Lett. A}, \textbf{26}, 1897, (2011).

\bibitem {wald84} R. M. Wald, "General Relativity", The University of Chicago Press, Chicago, 1984.

\bibitem{jaffe05} T.R. Jaffe, S. Hervik, A.J. Banday, K.M. Gorski, "On the viability of Bianchi type VII-H models with dark energy, \textit{astro-ph/0512433v1}, (2005). 

\bibitem{planck15}  P.A.R. Ade et al, "Planck 2015 results. XIII. Cosmological Parameters", \textit{arXiv: 1502.01589v3}, (2016).

\bibitem{pont07} A. Pontzen, A. Challinor, " Bianchi model CMB polarization and its implications for CMB anomalies", \textit{Mon. Not, Roy. Astron. Soc.}, \textbf{380}, 1387, (2007).

\bibitem{tripathy10} S.K. Tripathy, D. Behera, T.R. Rotray, "Anisotropic universe with cosmic string and bulk viscosity", \textit{Astrophys. Space Sci.}, \textbf{325}, 93, (2010).

\bibitem{brevik} I. Brevik, "Viscous Induced cosmic in the phantom barrier", \textit{Entropy}, \textbf{17}, 6318, (2015).

\bibitem{J} P. A. R. Ade et al., "Planck Collaboration XXIII, Planck 2013 results. XXIII, Isotropy and statistics of the CMB", \textit{Astronomy Astrophys}, \textbf{571}, A23, (2014).

\bibitem{K} P. A. R. Ade et al., "Planck Collaboration XVI, Planck 2015 results. XVI, Isotropy and statistics of the CMB", \textit{Astronomy Astrophys}, \textbf{594}, A16, (2016).

\bibitem{mishra2} B. Mishra, S. K. Tripathy, "Anisotropic dark energy model with a hybrid scale factor", \textit{Modern Physics Letter A},\textbf{30}, 1550175, (2015).

\bibitem{mishra3} B. Mishra, P K Sahoo, S K Tripathy, "Pressure anisotropy and dark energy models in scale invariant theory of gravitation" \textit{Astrophys. Space Sci.}, \textbf{356}, 163, (2015).

\bibitem{suresh11} S.Kumar, A.K. Yadav, "Some Bianchi type-V models of accelerating universe with dark energy", \textit{Mod. Phys. Lett A}, \textbf{26}, 647, (2011).

\bibitem{H} P. A. R. Ade et al., "Planck 2013 results. XVI. Cosmological parameters", \textit{Astronomy Astrophys}, \textbf{571}, A16 (2014).

\bibitem{I} M Betoule et. al., "Improved cosmological constraints from a joint analysis of the SDSS-II and SNLS supernova samples" \textit{Astronomy Astrophys}, \textbf{568}, A22, (2014).


\end{thebibliography}
\end{document}